\documentclass[12pt,letterpaper]{article}

\usepackage{natbib,hyperref}

\usepackage[ left=1in, top=1in, right=1in, bottom=1in]{geometry}
\usepackage{graphicx,bm,colonequals,amsmath,amssymb,bbold,url,xcolor,bbm}
\usepackage{array,tabularx,multirow}
\usepackage{enumitem}
\usepackage[font={footnotesize}]{caption,subcaption}
\usepackage[utf8]{inputenc}
\usepackage{enumitem}
\usepackage{soul} 
\usepackage{placeins} 
\usepackage[normalem]{ulem}

\usepackage[ruled,vlined]{algorithm2e}
\SetKwInput{KwInput}{Input}
\usepackage{algorithmic}
\usepackage{array,framed}
\usepackage{float}


\graphicspath{{plots/}}

\usepackage{mathtools}
\mathtoolsset{showonlyrefs} 

\bibpunct[, ]{(}{)}{;}{a}{,}{,}
   
\usepackage{amsthm}
\newtheoremstyle{propstyle} 
    {2mm}                    
    {1mm}                    
    {\itshape}                   
    {}                           
    {\scshape}                   
    {.}                          
    {.5em}                       
    {}  
\theoremstyle{propstyle}

\theoremstyle{propstyle}

\theoremstyle{propstyle}

\theoremstyle{propstyle}
\newtheorem{claim}{Claim}
\theoremstyle{propstyle}

\usepackage{array,framed}

\makeatletter
\renewcommand{\paragraph}{%
  \@startsection{paragraph}{4}%
  {\z@}{2ex \@plus 1ex \@minus .2ex}{-1em}%
  {\normalfont\normalsize\bfseries}%
}
\makeatother

\DeclareMathAlphabet\mathbfcal{OMS}{cmsy}{b}{n}

\newcommand{\bv}{\mathbf{v}}

\newcommand{\bs}{\mathbf{s}}

\newcommand{\bx}{\mathbf{x}}
\newcommand{\by}{\mathbf{y}}

\newcommand{\bw}{\mathbf{w}}
\newcommand{\bS}{\mathbf{S}}

\newcommand{\bP}{\mathbf{P}}
\newcommand{\bA}{\mathbf{A}}

\newcommand{\bJ}{\mathbf{J}}

\newcommand{\bE}{\mathbf{E}}
\newcommand{\bL}{\mathbf{L}}

\newcommand{\bI}{\mathbf{I}}

\newcommand{\bH}{\mathbf{H}}
\newcommand{\bU}{\mathbf{U}}

\newcommand{\bK}{\mathbf{K}}

\newcommand{\bQ}{\mathbf{Q}}

\newcommand{\bR}{\mathbf{R}}

\newcommand{\bfzero}{\mathbf{0}}

\newcommand{\bfmu}{\bm{\mu}}

\newcommand{\bfepsilon}{\bm{\epsilon}}

\newcommand{\bfSigma}{\bm{\Sigma}}

\DeclareMathOperator*{\cov}{Cov}

\newcommand{\normal}{\mathcal{N}}

\DeclareFontFamily{OT1}{pzc}{}
\DeclareFontShape{OT1}{pzc}{m}{it}{<-> s * [1.2] pzcmi7t}{}
\DeclareMathAlphabet{\mathpzc}{OT1}{pzc}{m}{it}

\newcommand{\nG}{n_{\mathcal{G}}}
\newcommand{\nT}{n_t}

\newcommand{\il}{{i_1,\ldots,i_l}}

\newcommand{\jm}{{j_1,\ldots,j_m}}

\newcommand{\jl}{{j_1,\ldots,j_\ell}}

\newcommand{\levol}{\mathbf{E}}

\DeclareMathOperator*{\chol}{chol}

\newcommand{\spM}{\bS}

\title{Scalable Spatio-Temporal Smoothing via Hierarchical Sparse Cholesky Decomposition}

\author{Marcin Jurek\thanks{Department of Statistics, Texas A\&M University} \and Matthias Katzfuss\footnotemark[1] \thanks{Corresponding author: \texttt{katzfuss@gmail.com}}}
\date{}

\begin{document}

\maketitle

\begin{abstract}
We propose an approximation to the forward-filter-backward-sampler (FFBS) algorithm for large-scale spatio-temporal smoothing. FFBS is commonly used in Bayesian statistics when working with linear Gaussian state-space models, but it requires inverting covariance matrices which have the size of the latent state vector. The computational burden associated with this operation effectively prohibits its applications in high-dimensional settings. We propose a scalable spatio-temporal FFBS approach based on the hierarchical Vecchia approximation of Gaussian processes, which has been previously successfully used in spatial statistics. On simulated and real data, our approach outperformed a low-rank FFBS approximation.
\end{abstract}

{\small\noindent\textbf{Keywords:} state-space model, spatio-temporal statistics, data assimilation, Vecchia approximation, smoothing}

\section{Introduction\label{sec:intro}}

Developments in data collection and storage technologies over the past decade have led to an unprecedented influx of data across scientific disciplines. Environmental sciences in particular have profited immensely from these advances. For example, frequent and high-resolution measurements of carbon dioxide acquired by the Orbiting Carbon Observatory \citep{oco2} helped to increase the understanding of CO$_2$ sinks and sources. Massive remotely-sensed data was demonstrated to be of help in determining the concentration of volcanic ash in the atmosphere \citep{bugliaro2021combining}, which is crucial for air traffic control and weather forecasting. Not all big data sets are collected using satellites however. Recently, Argo, a large system of autonomous floats, was deployed worldwide to collect data used in studying ocean temperature changes and the water cycle \citep{jayne2017argo}. 

Data sets of this kind are often spatio-temporal in nature and typically measure some scientifically interesting phenomenon. This leads the researchers to analyze them using a ``mechanistic'' approach. Within this paradigm,changes in time are represented by a (possibly discretized) differential equation, while the residual variation in space is captured using a purely statistical model \citep[e.g.,][]{wikle2019spatio}. Using this framework, data can be used to estimate the true value of the variable of interest, filling in the gaps where the observations are missing or inaccurate due to measurement errors, as well as to infer the unknown parameters. The first of these objectives is traditionally accomplished using the Kalman filter \citep{Kalman1960} and smoother \citep[also known as the Rauch–Tung–Striebel smoother,][]{rauch1965maximum}, while parameter inference is possible using a Gibbs sampler, often based on the forward-filter-backward-sampler \citep[FFBS;][]{Durbin2002, Fruhwirth1994, Carter1994}.

A major challenge in using these existing techniques with big environmental data is their poor scalability as the number of observations or grid points grows. Specifically, the computational cost of the canonical versions of filtering and smoothing methods is cubic in the number of observations at each time point. Countless approximations have been developed to address these problems, many of them being focused on filtering inference \citep[see, e.g.,][and the citations therein]{Jurek2022}. A particularly promising class of methods, which have recently gained prominence, are algorithms using an ensemble to represent the distribution of the state vector \citep{evensen2022data, grudzien2021fast}, most notably the Ensemble Kalman Filter \citep[e.g.][]{evensen1994sequential, katzfuss2020ensemble}. Variational approaches led to the development of the so-called 4D-VAR algorithm \citep[see e.g.][for a comprehensive introduction]{evensen2022data}, which has found mission-critical operational applications \citep[see e.g.][]{ECMWFdocs}.

Relatively little attention has been devoted to smoothing. Among the existing works, \citet{Katzfuss2012} propose a method based on a low-rank approximation of the latent Gaussian random field, which scales well but may not be able to reproduce fine-scale features. \citet{stroud2010ensemble} suffers from somewhat of the opposite problem, because it relies on tapering the sample covariance matrix and thus may struggle with smooth covariance functions \citep[see numerical experiments in][]{jurek2021multi}. \citet{sigrist2015stochastic} propose an approach based on spectral methods which are limited to observations on a regular grid. Another technique for approximate smoothing inference uses particle-based methods \citep{carvalho2010particle}, but such methods cannot be used when the dimension of the latent space exceeds several hundred because of particle collapse. A method that is perhaps the closest to our in spirit is based on the ensemble Kalman smoother, which is reviewed and extended in  \citet{katzfuss2020ensemble}. However, it also requires additional approximations such as tapering, and the number of distinct samples that it produces is always equal to the size of the ensemble, which can be inefficient.

We propose a scalable algorithm for generating samples from the smoothing distribution, directly approximating the FFBS algorithm, based on the hierarchical Vecchia approximation that has previously been used for spatio-temporal filtering \citep{Jurek2022}. We summarize the previous results developed in the context of filtering and extend them to approximate smoothing inference. This is not straightforward because matrix approximations used in previous work cannot be easily applied in the context of smoothing. We conducted numerical experiments showing that our sampler outperformed a low-rank approximation and showing how our method can be used to estimate unknown parameters using a Gibbs sampler. We also applied our method to a real data set and showed that it performed better than a competing approach. The code and data needed to reproduce our results can be found at \url{https://github.com/marcinjurek/scalable-FFBS}.

This paper is organized as follows. Section \ref{sec:SSM} introduces notation and briefly describes the linear Gaussian state space model and the canonical methods used for filtering, smoothing and sampling. Section \ref{sec:SCF} presents sparse Cholesky factorization and the hierarchical Vecchia approximation. In Section \ref{sec:approximate-methods} we propose approximations to the canonical methods from Section \ref{sec:SSM} and conclude with a scalable version of the FFBS algorithm. Section \ref{sec:comparison} contains numerical experiments which demonstrate excellent performance of our approximate methods. Section \ref{sec:tpw} discusses an application to a real data set. Section \ref{sec:conclusion} concludes and proposes directions for future research.

\section{Spatio-temporal state-space model} \label{sec:SSM}
Consider a Gaussian process $x(\cdot)$ defined over a domain $[1, 2, \dots, T] \times \mathcal{D} \subset \mathbb{R}^2$. Let $\mathcal{S} = \{\bs_1, \bs_2, \dots, \bs_{\nG}\}$ be a grid over $\mathcal{D}$ and let $\bx_t = \left[x(t, \bs_1), x(t, \bs_2), \dots, x(t, \bs_{\nG})\right]^\top$. Note that the grid is taken to be the same at all time points, which is common in the case of big environmental data sets, for example those collected using remote sensing.
We assume that the dynamics of the process at the subsequent time points can be expressed as an autoregressive model:
\begin{equation}
    \bx_t = \levol_t \bx_{t-1} + \bw_t, \quad \bw_t \sim \normal_{\nG}(\bfzero, \bQ_t), \label{eq:state-model}
\end{equation}
where the evolution matrix $\levol_t$ is assumed to be sparse. We do not make any special additional assumptions regarding the covariance matrix $\bQ_t$. The initial state follows a normal distribution: $\bx_0 \sim \normal_{\nG}(\bfmu_{0|0}, \bfSigma_{0|0})$. 

We consider a situation in which at each time point we are given $\by_t$, an $\nT$-dimensional vector of data observed at time $t=1,2,\ldots,T$, related to the true process through a linear function:
\begin{equation}
    \by_t = \bH_t \bx_t + \bv_t,\quad \bv_t \sim \normal_{\nT}(\bfzero, \bR_t), \label{eq:data-model}
\end{equation}
We assume that observation error covariance matrix $\bR_t$ is diagonal. (This can be extended to block-diagonal $\bR_t$ with small blocks.) We use $\by_{1:t} \colonequals (\by_1^\top, \ldots, \by_t^\top)^\top$ to denote a vector of observations from time $1$ to time $t$ and we define $\bx_{1:t}$ analogously. At each time $t$, the locations of the observations $\by_t$ can be a (different) subset of size $n_t$ of the grid $\mathcal{S}$, indicated by the $n_t \times n$ matrix $\bH_t$. In this paper we are interested in obtaining the filtering and smoothing distributions of $\bx_t$ for $t=1, 2, \dots, T$, i.e. $p(\bx_t|\by_{1:t})$ and $p(\bx_t|\by_{1:T})$, respectively. To accomplish this goal, we start with the canonical algorithms for filtering and generating samples from the smoothing distribution.

\subsection{The filtering distribution} 

Under the assumptions introduced in Section \ref{sec:SSM} the filtering distribution, $[\bx_t\mid \by_{1:t}]$ is Gaussian and can be obtained using the Kalman filter \citep{Kalman1960}. We use $\bfmu_{t|t}$ to denote $\mathbb{E}[\bx_t\mid \by_{1:t}]$ and set $\bfSigma_{t|t}\colonequals \cov(\bx_t\mid \by_{1:t})$. To derive the Kalman filtering procedure, we first give the one-step ahead forecasting distribution:
\begin{align} \label{eq:forecasting-dist}
    \bx_t\mid \by_{1:t-1} \sim \normal_{\nG}(\bfmu_{t|t-1}, \bfSigma_{t|t-1}),
\end{align}
where $\bfmu_{t|t-1} \colonequals \levol_t \bfmu_{t-1|t-1}$ and $\bfSigma_{t|t-1}\colonequals\levol_t\bfSigma_{t-1|t-1}\levol_t^\top + \bQ_t$.

Based on Bayes' theorem, it follows that $[\bx_t\mid \by_{1:t}] \propto [\by_t \mid \bx_t] [\bx_t\mid \by_{1:t-1}]$. Thus, we have 
\begin{align}
    \bfmu_{t|t} & \colonequals \bfmu_{t|t-1} + \bK_t(\by_t - \bH_t \bfmu_{t|t-1}), \\
    \bfSigma_{t|t} & \colonequals\bfSigma_{t|t-1} - \bK_t \bH_t \bfSigma_{t|t-1},
\end{align}
where $\bK_t\colonequals\bfSigma_{t|t-1} \bH_t^\top (\bH_t \bfSigma_{t|t-1} \bH^\top + \bR_t)^{-1}$ is the $\nG\times \nT$ Kalman gain matrix.

\begin{algorithm}
\caption{\label{alg:KF}Kalman Filter (KF)}
\KwInput{ moments of the initial distribution $\bfmu_{0|0}, \bfSigma_{0|0}$, evolution model $\left\{ \bE_t, \bQ_t\right\}_{t=0}^T$, observation model $\left\{ \bH_t, \bR_t \right\}_{t=0}^T$, data $\left\{\by_t\right\}_{t=0}^T$ }
\KwResult{ moments of the filtering distribution $\left\{\bfmu_{t|t}, \bfSigma_{t|t}\right\}_{t=0}^T$ }
\begin{algorithmic}[1] 
\FOR{$t=1, 2, \ldots, T$}
    \STATE Compute forecast mean $\bfmu_{t|t-1} = \bE_t \bfmu_{t-1|t-1}$
    \STATE Compute forecast covariance $\bfSigma_{t|t-1} = \bE_t \bfSigma_{t-1|t-1} \bE_t^T + \bQ_t$.
    \STATE Calculate $\bK_t = \bfSigma_{t|t-1} \bH_t^\top (\bH_t \bfSigma_{t|t-1} \bH_t^\top + \bR_t)^{-1}$.
    \STATE Calculate filtering mean $\bfmu_{t|t} = \bfmu_{t|t-1} + \bK_t(\by_t - \bH_t \bfmu_{t|t-1})$.
    \STATE Calculate filtering covariance $\bfSigma_{t|t} = \bfSigma_{t|t-1} - \bK_t \bH_t \bfSigma_{t|t-1}$.
\ENDFOR
\end{algorithmic} 
\end{algorithm}

\subsection{Kalman smoother}

Computing the smoothing distribution can be accomplished using the Kalman smoother \citep{rauch1965maximum}. Let $\bfmu_{t|T} \colonequals \mathbb{E}(\bx_t\mid \by_{1:T})$ and $\bfSigma_{t|T} \colonequals \cov(\bx_t\mid \by_{1:T})$. Then the linear Gaussian state-space model of Section \ref{sec:SSM} implies that the smoothing distribution will also be Gaussian: $\bx_t \mid \by_{1:T} \sim \normal_{\nG}(\bfmu_{t|T}, \bfSigma_{t|T})$. Notice that $$[\bx_t\mid \by_{1:T}] = \int [\bx_t\mid \bx_{t+1}, \by_{1:T}] [\bx_{t+1} \mid \by_{1:T}]\, d\bx_{t+1},$$
where $[\bx_t\mid \bx_{t+1}, \by_{1:T}] \propto [\bx_{t+1} \mid \bx_t] [\bx_t\mid \by_{1:t}]$. It follows that the conditional mean and conditional covariance in the smoothing distribution are given by 
\begin{align}
    \bfmu_{t|T}&\colonequals \bfmu_{t|t} + \bJ_t(\bfmu_{t+1|T} - \bfmu_{t+1|t}), \label{eq:smoothing-mean}\\
    \bfSigma_{t|T} &\colonequals \bfSigma_{t|t} + \bJ_t(\bfSigma_{t+1|T} - \bfSigma_{t+1|t}) \bJ_t^\top,
\end{align}
where $\bJ_t\colonequals\bfSigma_{t|t}\levol_{t+1}^\top \bfSigma_{t+1|t}^{-1}$. 

\begin{algorithm}
\caption{\label{alg:KS}Kalman Smoother (KS)}
\KwInput{ moments of the initial distribution $\bfmu_{0|0}, \bfSigma_{0|0}$, evolution model $\left\{ \bE_t, \bQ_t\right\}_{t=0}^T$, observation model $\left\{ \bH_t, \bR_t \right\}_{t=0}^T$, data $\left\{\by_t\right\}_{t=0}^T$ }
\KwResult{ moments of the smoothing distribution $\left\{\bfmu_{t|T}, \bfSigma_{t|T}\right\}_{t=0}^T$ }
\begin{algorithmic}[1] 
\STATE Obtain moments of the filtering distribution $\left\{\bfmu_{t|t}, \bfSigma_{t|t}\right\}_{t=0}^T$ using KF (Algorithm \ref{alg:KF}).
\FOR{$t=T-1, T-2, \ldots, 1$}
    \STATE Compute $\bJ_t = \bfSigma_{t|t}\levol_{t+1}^\top \bfSigma_{t+1|t}^{-1}$
    \STATE Compute smoothing mean $\bfmu_{t|T} = \bfmu_{t|t} + \bJ_t(\bfmu_{t+1|T} - \bfmu_{t+1|t})$,
\ENDFOR
\end{algorithmic} 
\end{algorithm}
The full Kalman Smoother typically can compute also the smoothing covariance matrix $\bfSigma_{t|T} = \bfSigma_{t|t} + \bJ_t(\bfSigma_{t+1|T} - \bfSigma_{t+1|t}) \bJ_t^\top$. We skip this calculation in our Algorithm \ref{alg:KS}, as it is not necessary for the construction of the algorithm which samples from the smoothing distribution.

\subsection{Forward Filter Backward Sampler (FFBS)}

In Bayesian statistics instead of calculating the full smoothing distribution, it is often enough to be able to draw samples from $[\bx_t\mid \bx_{t+1}, \by_{1:T}]$. This is particularly true in Markov Chain Monte Carlo (MCMC) -based methods. Inspired by this fact, some authors \citep{Fruhwirth1994, Carter1994, Durbin2002} developed algorithms which draw a sample from \eqref{eq:data-model}-\eqref{eq:state-model} and then linearly transform it based on actual observations from \eqref{eq:data-model} to obtain a sample from the smoothing distribution. It is preferable to simulation using moments generated by the Kalman smoother, which would require, in general, factorization of all smoothing covariance matrices $\bfSigma_{t|T}$. We briefly summarize the algorithm known as forward filter backward sampler \cite{Durbin2002} below, using the helpful insights from \citet{Jarocinski2015}.

\begin{algorithm}
\caption{\label{alg:FFBS}Forward Filter Backward Sampler \citep{Durbin2002, Jarocinski2015}}
\KwInput{ moments of the initial distribution $\bfmu_{0|0}, \bfSigma_{0|0}$, evolution model $\left\{ \bE_t, \bQ_t\right\}_{t=0}^T$, observation model $\left\{ \bH_t, \bR_t \right\}_{t=0}^T$, data $\left\{\by_t\right\}_{t=0}^T$, desired number of samples $N_\text{samp}$ }
\KwResult{ sample from the smoothing distribution: $\bx_{1:T}$ }
\begin{algorithmic}[1] 

\STATE Generate $\hat{\bx}_{0|0} \sim \normal_{\nG}(\mathbf{0}, \bfSigma_{0|0})$.
\STATE Generate $\hat{\bx}_{1:T}$ and $\hat{\by}_{1:T}$ using \eqref{eq:data-model} - \eqref{eq:state-model}.
\STATE Calculate $\by_{1:T}^*$ where $\by_t^* = \by_t - \hat{\by}_t$.
\STATE Use KS (Algorithm \ref{alg:KS}) to obtain $\{\hat{\bfmu}_{t|T}\}_{t=1}^T$, where $\hat{\bfmu}_{t|T} = \mathbb{E}(\bx_{1:T}|\hat{\by}^*_{1:t})$.
\FOR{$t=1, \dots, T$}
\STATE $\bx_t = \hat{\bx}_t + \hat{\bfmu}_{t|T}$ is a sample from $\left[ \bx_t | \by_{1:T} \right]$.
\ENDFOR

\end{algorithmic} 
\end{algorithm}

We note, that a sample from the smoothing distribution can also be used as the approximation of the full distribution. For example if we are interested in prediction, the sample mean and quantiles can be used as a tool for making predictions and quantifying uncertainty, respectively.

\subsection{Computational complexity}

Algorithms \ref{alg:KS} - \ref{alg:FFBS} rely on calculating the correction factor $\bJ_t$, which requires computing the inverse of the forecast covariance matrix $\bfSigma_{t|t-1}$. In the case of Algorithm \ref{alg:KF}, a prerequisite for the other two, we also need to obtain the Kalman gain matrix $\bK_t$ which is a linear function of the inverse of $\bH_t \bfSigma_{t|t-1} \bH^\top + \bR_t$. This proves to be the computational bottleneck, since the number of operations required for matrix inversion is proportional to the cube of its dimension. As the size of the grid $\nG$ and the number of observations at each time point $\nT$ grow, these inversion operations take a prohibitive amount of time. 
In the next section we review the sparse Cholesky factorization method and subsequently show how it can be used to approximate Algorithms \ref{alg:KF} - \ref{alg:FFBS}. 

\section{Sparse Cholesky factorization}\label{sec:SCF}

\subsection{Hierarchical Vecchia (HV) approximation}

In this section we describe the hierarchical Vecchia (HV) approximation. It has recently been shown that this approach ensures that the sparsity of the approximate Cholesky factor of the filtering covariance matrix is the same at all time points \citep{Jurek2022}. Moreover, following the findings of \citet{schaefer2020} the approximation to the forecast distribution at each time point is optimal in the sense of KL-divergence, given the sparsity pattern . Here we summarize a special case of the Vecchia approximation which was shown to be near optimal \citep{Zilber2021} and which additionally has the property of preserving the sparsity of the Cholesky decomposition of the covariance matrix under inversion. As we show in the following sections, this characteristic is fundamental for a construction of a scalable FFBS. 

We start by defining an order relation $\prec$ among the elements of the grid $\mathcal{S}$ using the maxmin ordering \citep{schaefer2020}. From now on we assume that the elements of $\bx_0$ are sorted according to $\prec$. 
We then define a directed acyclic graph over the subsets of elements of $\bx_0$ in the following way. We begin by selecting the first $r_0$ elements of $\bx_0$, which we call knots, and label them as $\mathcal{K}^0$. Next we partition the remaining $\nG-r_0$ variables into $J$ groups $G_1, \dots, G_J$ and for each preserve the order $\prec$ truncated to members of that group. Finally, we select $r_1$ knots from each group and label them as $\mathcal{K}_j$ for $j=1, \dots, J$. Variables $\mathcal{K}^1 = \left\{\mathcal{K}_1, \dots, \mathcal{K}_J\right\}$ form the next level of the hierarchy. 

The remaining elements of each group are further partitioned. For example, the $\#G_j-r_1$ remaining elements of $G_j$ are divided into sets $G_{j,1}, \dots, G_{j,J}$. Then $r_2$ first elements from each of those smaller groups are put into sets $\mathcal{K}_{j,1}, \dots, \mathcal{K}_{j,J}$. In this way, we obtain the second level of the hierarchy $\mathcal{K}^2 = \left\{\mathcal{K}_{1,1}, \dots, \mathcal{K}_{1,J}, \mathcal{K}_{2,1}, \dots, \mathcal{K}_{2,J}, \dots, \mathcal{K}_{J,J}\right\}$.

\begin{figure}
    \centering
    \includegraphics[width=\textwidth]{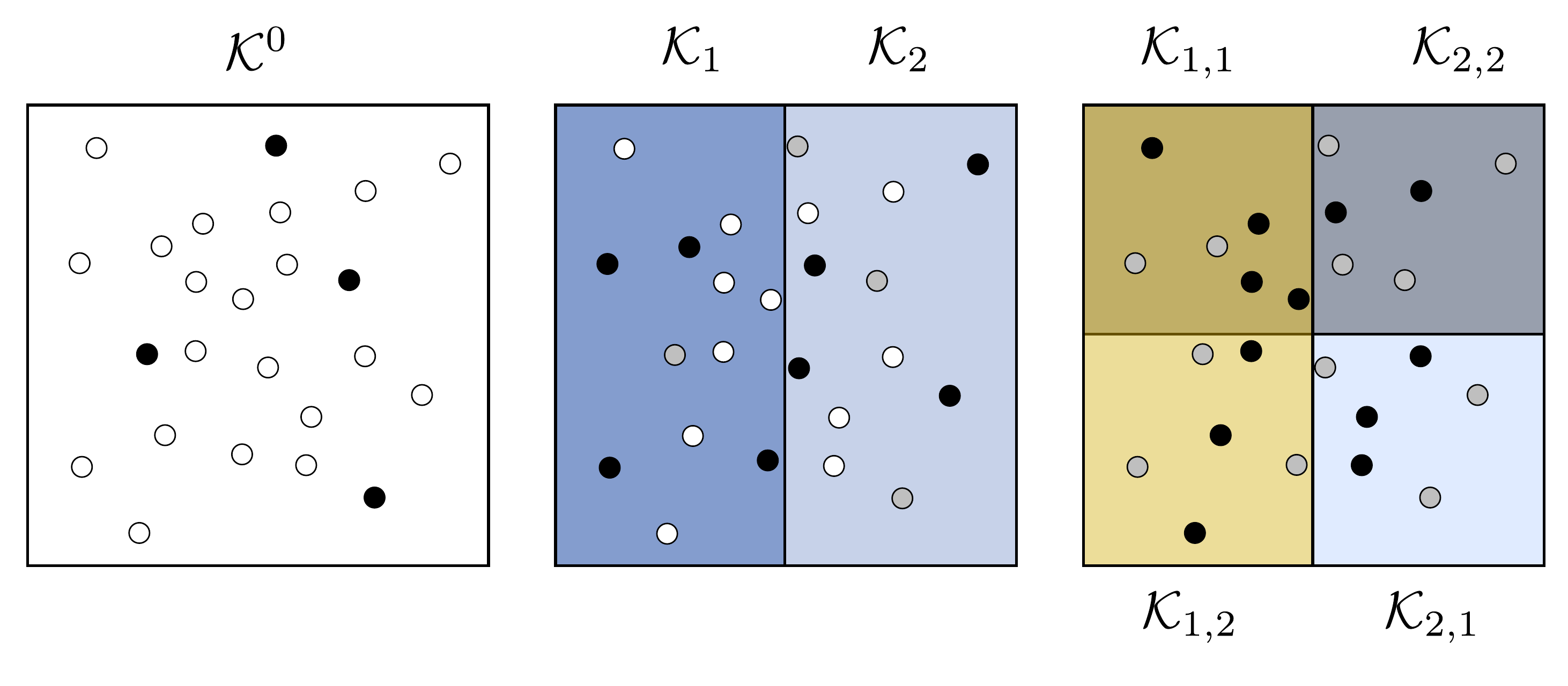}
    \caption{Construction of the hierarchical Vecchia approximation. The first panel shows the entire domain with each dot representing an element of $\bx_0$; note that they do not need to be regularly spaced. The black dots, in accordance with the label above, represent the elements selected as $\mathcal{K}^0$. The second panel shows the domain split in two. The elements of $\bx_0$ corresponding to locations in the left half are assigned to $G_1$, while the remaining elements are assigned to $G_2$. Black dots within each group denote elements of $\mathcal{K}^1$, the gray dots stand for elements already assigned to $\mathcal{K}^0$ and the remaining dots are white. The right panel shows another level of the hierarchy with each quadrant, from the top-left and counter-clockwise, the area covering $G_{1,1}, G_{1,2}, G_{2, 1}$ and $G_{2, 2}$. Similar to the middle panel, the grey dots represent elements of $\mathcal{K}^0 \cup \mathcal{K}^1$ and black dots represent elements of $\mathcal{K}^2$ split into four sets $K_{i,j}$ such that $K_{i,j}\subset G_{i,j}$ for $i,j \in \{1, 2\}$.}
    \label{fig:vecchia-approx}
\end{figure}

This hierarchy can be visually represented in the form of a directed graph $\mathcal{G} = (V, E)$ where $V = \mathcal{K}^0 \cup \mathcal{K}^1 \cup \dots$ and $E$ is defined as follows. For two vertices $\mathcal{K}_\jm$ and $\mathcal{K}_\il$ we have $\mathcal{K}_\jm \rightarrow \mathcal{K}_\il$ if $\mathcal{K}_\il \subset G_\jm$ and $\mathcal{K}_\jm \leftarrow \mathcal{K}_\il$ if $\mathcal{K}_\jm \subset G_\jl$. The construction of this hierarchy is illustrated in Figure \ref{fig:vecchia-approx}.

We also introduce lexicographic order $\prec_L$ on vertices $\mathcal{K}_\jm$ with respect to their subscripts and define $\bS$ to be an adjacency matrix of graph $\mathcal{G}$. Note that this matrix is lower triangular because for $w, v \in V$ we can have $w \rightarrow v$ only if $w \prec_L v$.

Further details of the HV construction can be found in \citet{Jurek2022}.

\subsection{Sparse Cholesky decomposition based on HV}

With the sparsity pattern encoding the HV approximation we now modify the standard Cholesky factorization algorithm in the following way. If an $i,j$-th element of the sparsity pattern matrix $\spM$ equals 1, we calculate the corresponding element of the Cholesky factor, using the the regular formula and set it to zero otherwise. Note also that given the HV construction the diagonal elements will always be calculated. Our approach is summarized in Algorithm \ref{alg:HCF}.

\begin{algorithm}[ht]
\caption{\label{alg:HCF} Hierarchical Cholesky factorization (HCF) }
\KwInput{ Sparsity pattern matrix $\spM$, p.d. matrix $\bA$ of size $n \times n$ }
\KwResult{ Sparse Cholesky factor $\bL$ }
\begin{algorithmic}[1]
    \FOR{ $i = 1, \dots, n$ }
        \FOR{ $j = 1, \dots i$ }
        \STATE $\bL_{i,j} = \spM_{i,j} \cdot (\bA_{i,j} - \sum_{k=1}^{j-1}\bL_{i,k}\bL_{j,k})/\bL_{j,j}$
        \ENDFOR
        \STATE $\bL_{i,i} = (\bA_{i,i} - \sum_{k=1}^{i-1}\bL_{i,k}^2)^{1/2}$
    \ENDFOR
\end{algorithmic}
\end{algorithm}

If we use $N$ to denote the maximum number of nonzero elements in a row of $\bS$, then the complexity of Algorithm \ref{alg:HCF} is $\mathcal{O}(nN^2)$. This is because line 3 requires $\mathcal{O}(N)$ operations and is executed at most $N$ times for each of the $n$ rows.

\section{Fast sampling using sparse Cholesky factorization}\label{sec:approximate-methods}

In this section, we show how Algorithm \ref{alg:HCF} (HCF) can be used to ensure the scalability of Algorithm \ref{alg:FFBS} (FFBS). Recall that the most computationally-instensive steps in Algorithm \ref{alg:FFBS} were those calculating the $\bK_t$ matrix in the forward pass and inverting the forecast covariance in the backward pass. We show how HCF can be used to accelerate both.

\subsection{Approximate filtering} \label{sec:HVF}

The application of hierarchical Cholesky factorization to filtering was described previously \citep{Jurek2022} and we briefly summarize it here. 
Unlike in Algorithm \ref{alg:KF} we do not calculate the entire filtering and forecast covariance matrices, $\bfSigma_{t|t-1}$ and $\bfSigma_{t|t}$, respectively, but rather their hierachical Cholesky factor. 
In particular, given the prescribed sparsity $\spM$, we approximate ${\bfSigma}_{t|t-1} \approx \widetilde{\bfSigma}_{t|t-1} = \bL_{t|t-1}\bL_{t|t-1}^\top$, where $\bL_{t|t-1} = \text{HCF}(\spM, \bfSigma_{t|t-1})$, which is optimal in the sense of KL divergence \citep{schaefer2020}. The computational benefits of using this approximation can be further taken advantage of \citep[][Section 3.3]{Jurek2022} as shown in the following

\begin{claim}\label{clm:L2U}
Assume $\bL_{t|t-1} = \text{\emph{HCF}}(\spM, \bfSigma_{t|t-1})$, where $\spM$ encodes the hierarchical Vecchia approximation, $\bfSigma_{t|t-1}$ is a (approximate or exact) forecast covariance matrix and that $\bP$ is an order reversing permutation matrix. We have 
$$
\bU_{t|t} = \bP\chol(\bP(\bL_{t|t-1}^{-\top}\bL_{t|t-1}^{-1} + \bH_t\bR_t^{-\top}\bH_t)\bP)\bP
$$
and
$$
\widetilde{\bfSigma}_{t|t} = \bU_{t|t}^{-\top}\bU_{t|t}^{-1}
$$
\end{claim}
We can thus define $\bL_{t|t} \colonequals \bU_{t|t}^{-\top}$. Then, as \citet{Jurek2022} noted, given $\bL_{t|t-1}$ with at most $N$ nonzero elements in a row, $\bL_{t|t}$ has the same sparsity pattern as $\bL_{t|t-1}$ and can be calculated in $\mathcal{O}(nN^2)$ time. These properties allow us to approximate Algorithm \ref{alg:KF} (Kalman Filter) using Algorithm \ref{alg:HVF}, which \citet{Jurek2022} show to have $\mathcal{O}(nN^2T)$ time complexity.

\begin{algorithm}
\caption{\label{alg:HVF} Hierarchical Vecchia filter (HVF)}
\KwInput{ moments of the initial distribution $\bfmu_{0|0}, \bfSigma_{0|0}$, evolution model $\left\{ \bE_t, \bQ_t\right\}_{t=0}^T$, observation model $\left\{ \bH_t, \bR_t \right\}_{t=0}^T$, data $\left\{\by_t\right\}_{t=0}^T$ and sparsity matrix $\bS$ }
\KwResult{ approximate representation of the filtering and forecast distributions $\left\{ \widetilde{\bfmu}_{t|t-1}, \bL_{t|t-1} \right\}_{t=0}^T$, $\left\{ \widetilde{\bfmu}_{t|t}, \bL_{t|t} \right\}_{t=0}^T$ }
\begin{algorithmic}[1] 

\STATE Calculate the HV sparsity matrix $\spM$ 
\STATE Calculate $\bL_{0|0} = \text{HCF}(\spM, \bfSigma_{0|0})$

\FOR{$t=1, 2, \ldots, T$}
    \STATE Compute $\widetilde{\bfmu}_{t|t-1}=\levol_t \widetilde{\bfmu}_{t-1|t-1}$
    \STATE Calculate the $(i,j)$-th elements of $\widetilde{\bfSigma}_{t|t-1} = \bE_t \bL_{t-1|t-1} \bL_{t-1|t-1}^\top\bE_t^\top + \bQ_t$, for $(i,j)$ such that $\spM_{i,j} = 1$
    \STATE Calculate the HCF of the forecast matrix $\bL_{t|t-1} = \text{HCF}(\spM, \widetilde{\bfSigma}_{t|t-1})$
    \STATE Calculate $\bL_{t|t}$ using Claim \ref{clm:L2U}.
    \STATE Compute $\widetilde{\bfmu}_{t|t} = \widetilde{\bfmu}_{t|t-1} + \bL_{t|t}\bL_{t|t}^\top\bH_t^{\top}\bR_t^{-1}\left(\by_t - \bH_t \widetilde{\bfmu}_{t|t-1}\right)$
\ENDFOR
\end{algorithmic} 
\end{algorithm}
Note that the approximate filtering and forecast means are denoted with a tilde over each symbol, to differentiate them from their exact counterparts calculated in Algorithm \ref{alg:KF}.

\subsection{Approximate sampling}\label{sec:HVS}

Following Algorithm \ref{alg:KS} we see that the most time consuming part of the backward pass is matrix inversion in line 3. Additionally, the multiplication of dense $\nG \times \nG$ matrices also requires much computation time for large $\nG$. These bottlenecks can be eliminated if matrices $\bfSigma_{t+1|t}$ and $\bfSigma_{t|t}$ are replaced with their hierarchical Cholesky factors $\bL_{t+1|t}$ and $\bL_{t|t}$, respectively. This substitution also decreases the cost of matrix multiplication, since all matrices in line 3 are now sparse. This let allows us to approximate Algorithm \ref{alg:FFBS} by proposing a scalable FFBS in Algorithm \ref{alg:sFFBS}.

\begin{algorithm}
\caption{\label{alg:HVS}Hierarchical Vecchia Smoother (HVS)}
\KwInput{ moments of the initial distribution $\bfmu_{0|0}, \bfSigma_{0|0}$, evolution model $\left\{ \bE_t, \bQ_t\right\}_{t=0}^T$, observation model $\left\{ \bH_t, \bR_t \right\}_{t=0}^T$, data $\left\{\by_t\right\}_{t=0}^T$ }
\KwResult{ approximate mean of the smoothing distribution $\left\{\widetilde{\bfmu}_{t|T}\right\}_{t=0}^T$ }
\begin{algorithmic}[1] 
\STATE Obtain representation of the forecast and filtering distributions $\left\{ \widetilde{\bfmu}_{t|t}, \bL_{t|t} \right\}_{t=0}^T$, $\left\{ \widetilde{\bfmu}_{t|t-1}, \bL_{t|t-1} \right\}_{t=0}^T$ using HVF (Algorithm \ref{alg:HVF})
\FOR{$t=T-1, T-2, \ldots, 1$}
    \STATE Compute approximate smoothing mean as $\widetilde{\bfmu}_{t|T} = \widetilde{\bfmu}_{t|t} + \bL_{t|t}\bL_{t|t}^\top \levol_{t+1}^\top \bL_{t+1|t}^{-\top}\bL_{t+1|t}^{-1}(\widetilde{\bfmu}_{t+1|T} - \widetilde{\bfmu}_{t+1|t})$,
\ENDFOR
\end{algorithmic} 
\end{algorithm}

Similar to Algorithm \ref{alg:HVF} we used symbols with a tilde to denote the approximations of corresponding variables in Algorithm \ref{alg:KS}. Regarding complexity of Algorithm \ref{alg:HVS}, $\bU_{t|t+1}$ is sparse with a known sparsity pattern which means that line 3 and can be executed in $\mathcal{O}(nN^2T)$ time \citep{Jurek2022}. Line 4 can be executed efficiently series of matrix-vector multiplications is performed instead. The matrices $\bL_{t|t}$ and $\bU_{t+1|t}$ have at most $N$ nonzero elements in each row \citep{jurek2021multi}. Therefore, if we recall the complexity of Algorithm \ref{alg:HVF} discussed in Section \ref{sec:HVF} and assume $\bE_t$ is sparse, then operations in line 4 have complexity $\mathcal{O}(nN)$. Consequently, Algorithm \ref{alg:HVS} can be executed in $\mathcal{O}(nN^2T)$ time.

\subsection{Scalable FFBS}

Using the approximations described in Sections \ref{sec:HVF} and \ref{sec:HVS} we can now provide an algorithm for a scalable FFBS. Following the approach adopted earlier in this section \ref{alg:FFBS} we used the tilde notation to indicate approximations. Notice that we use Vecchia approximation in order to quickly calculate the square roots $\{\bL_t\}_{t=1}^T$ of the model error covariance matrices $\{\bQ_t\}_{t=1}^T$. These square roots are then be used for quick generation the synthetic data.

\begin{algorithm}
\caption{\label{alg:sFFBS}Scalable FFBS}
\KwInput{ moments of the initial distribution $\bfmu_{0|0}, \bfSigma_{0|0}$, evolution model $\left\{ \bE_t, \bQ_t\right\}_{t=0}^T$, observation model $\left\{ \bH_t, \bR_t \right\}_{t=0}^T$, data $\left\{\by_t\right\}_{t=0}^T$, and sparsity pattern $\bS$ }
\KwResult{ sample $\bx_{1:T}$ from the approximate smoothing distribution}
\begin{algorithmic}[1] 
\STATE Calculate $\bL_{0|0} = \text{HCF}(\bS, \bfSigma_{0|0})$.
\STATE Generate $\bfepsilon_0 \sim \normal_{\nG}(\mathbf{0}, \bI_{\nG})$, and set $\hat{\bx}_{0|0} = \bL_{0|0}\bfepsilon_0$. 
\FOR{$t=1, \ldots, T-1$}
    \STATE Calculate $\bL^Q_t = \text{HCF}(\bS, \bQ_t)$
    \STATE Calculate $\bfepsilon_t \sim \normal_{\nG}(\mathbf{0}, \bI_{\nG})$ and set $\hat{\bw}_t = \bL^Q_t \bfepsilon_t.$
\ENDFOR

\STATE Generate $\hat{\bx}_{1:T}$ and $\hat{\by}_{1:T}$ using \eqref{eq:data-model} - \eqref{eq:state-model}, replacing $\bw_t$ with $\hat{\bw}_t$.
\STATE Calculate $\by_{1:T}^*$ where $\by_t^* = \by_t - \hat{\by}_t$.
\STATE Use HVS (Algorithm \ref{alg:KS}) to obtain $\{\hat{\widetilde{\bfmu}}_{t|T}\}_{t=1}^T$, where $\hat{\widetilde{\bfmu}}_{t|T} = \mathbb{E}(\bx_t|\hat{\by}^*_{1:t})$.
\FOR{t = 1, dots, T}
\STATE $\bx_t = \hat{\bx}_t + \hat{\widetilde{\bfmu}}_{t|T}$ is a sample from an approximation of $\left[ \bx_t | \by_{1:T} \right]$.
\ENDFOR
\end{algorithmic} 
\end{algorithm}

\subsection{Computational complexity}\label{sec:complexity}

Using the Hierarchical Vecchia approximation substantially reduces the computational cost of sampling from the smoothing distribution. If $\bS$ corresponds to a hierarchical Vecchia approximation, the first line of the algorithm can be calculated in $\mathcal{O}(nN^2)$ time. The computationally intense operation in the second line is the matrix-vector multiplication, but because $\bL_{0|0}$ has the same sparsity pattern as $\bS$, this product can be obtained in $\mathcal{O}(nN)$ time. Analogous arguments let us conclude that the total cost of line 3 is $\mathcal{O}(nN^2T)$. Generating synthetic data $\hat{\bx}_{1:T}$ and $\hat{\by}_{1:T}$ can be done in $\mathcal{O}(nNT)$ time, because we assumed that the evolution matrix $\bE_t$ is sparse and that $\bR_t$ is block diagonal with small blocks. The only operation in the remaining lines is the use of hierarchical Vecchia smoother in line 9, which requires $\mathcal{O}(nN^2T)$ time.

A typical user of Algorithm \ref{alg:sFFBS} will typically generate $N_{\text{samp}} > 1$ samples from the approximate smoothing distribution, which means that it will take $\mathcal{O}(nN^2TN_{\text{samp}})$ time.




\section{Numerical comparison\label{sec:comparison}}

\subsection{Setup}\label{sec:comparison-setup}
In this section we evaluate our scalable FFBS using simulated data. 

\begin{figure}[ht]
    \centering
    \begin{subfigure}[b]{0.45\textwidth}
        \includegraphics[trim={30 30 30 30},clip,width=1.0\textwidth]{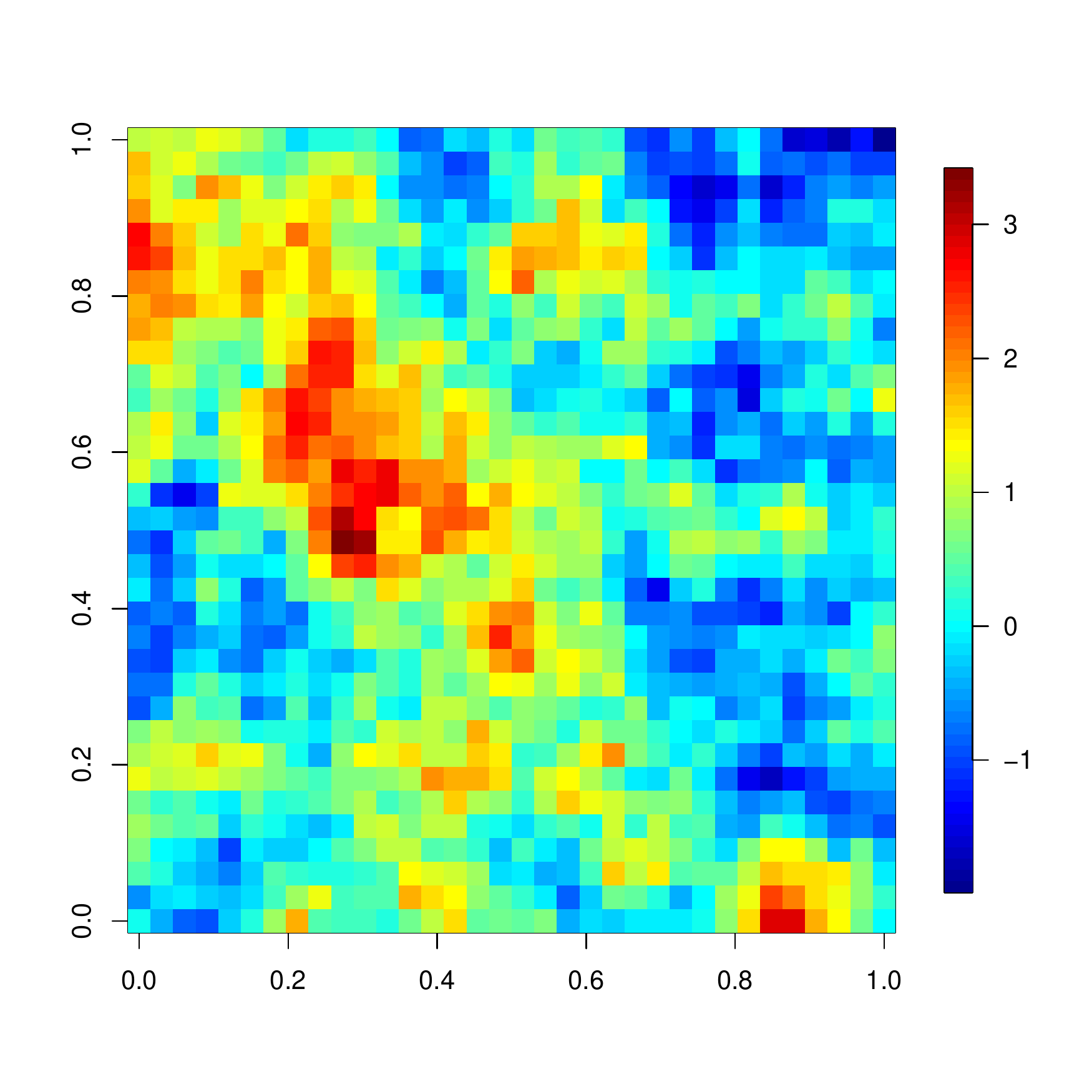}
        \caption{t=1}
    \end{subfigure}
    \begin{subfigure}[b]{0.45\textwidth}
        \includegraphics[trim={30 30 30 30},clip,width=1.0\textwidth]{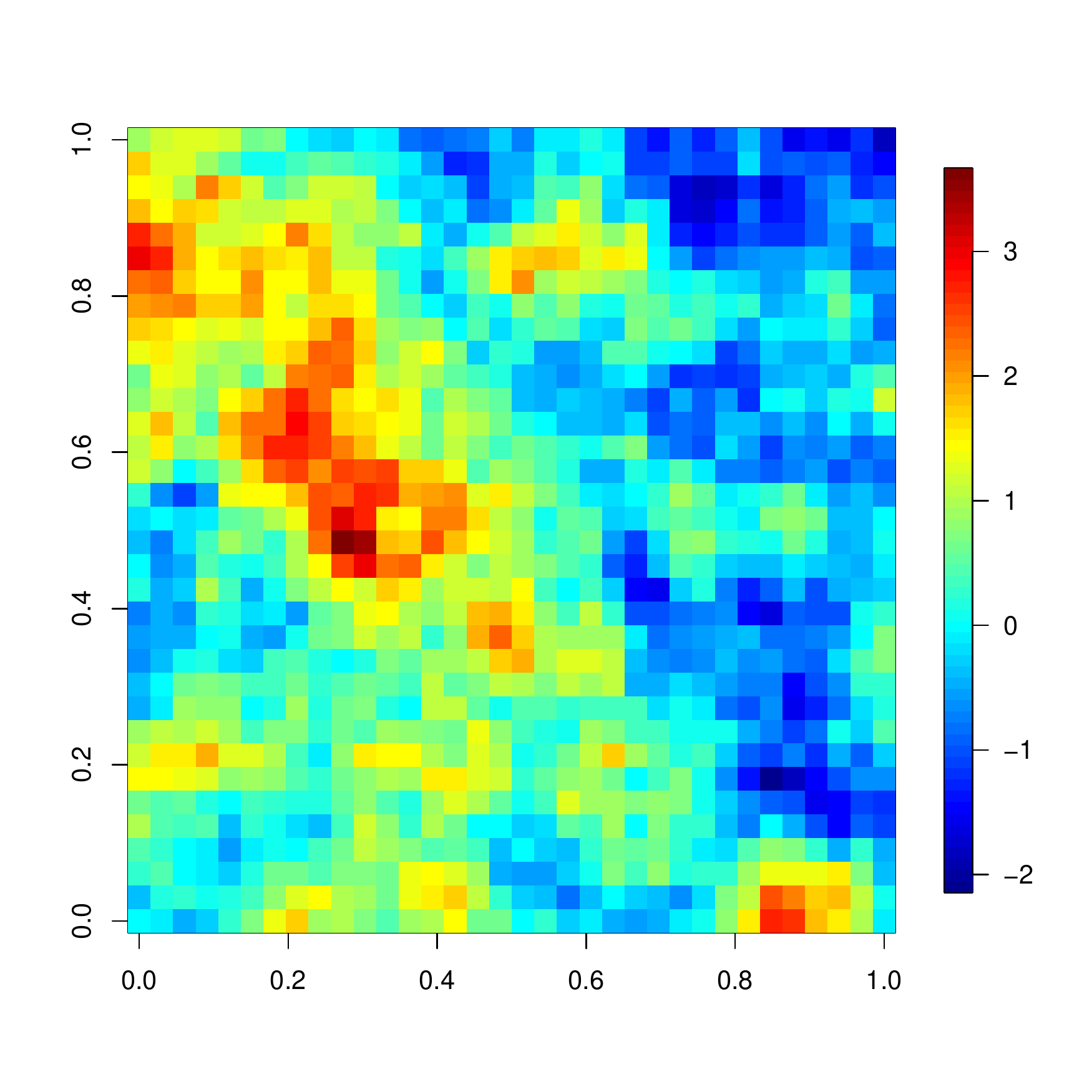}
        \caption{t=2}
    \end{subfigure}
    \begin{subfigure}[b]{0.45\textwidth}
        \includegraphics[trim={30 30 30 30},clip,width=1.0\textwidth]{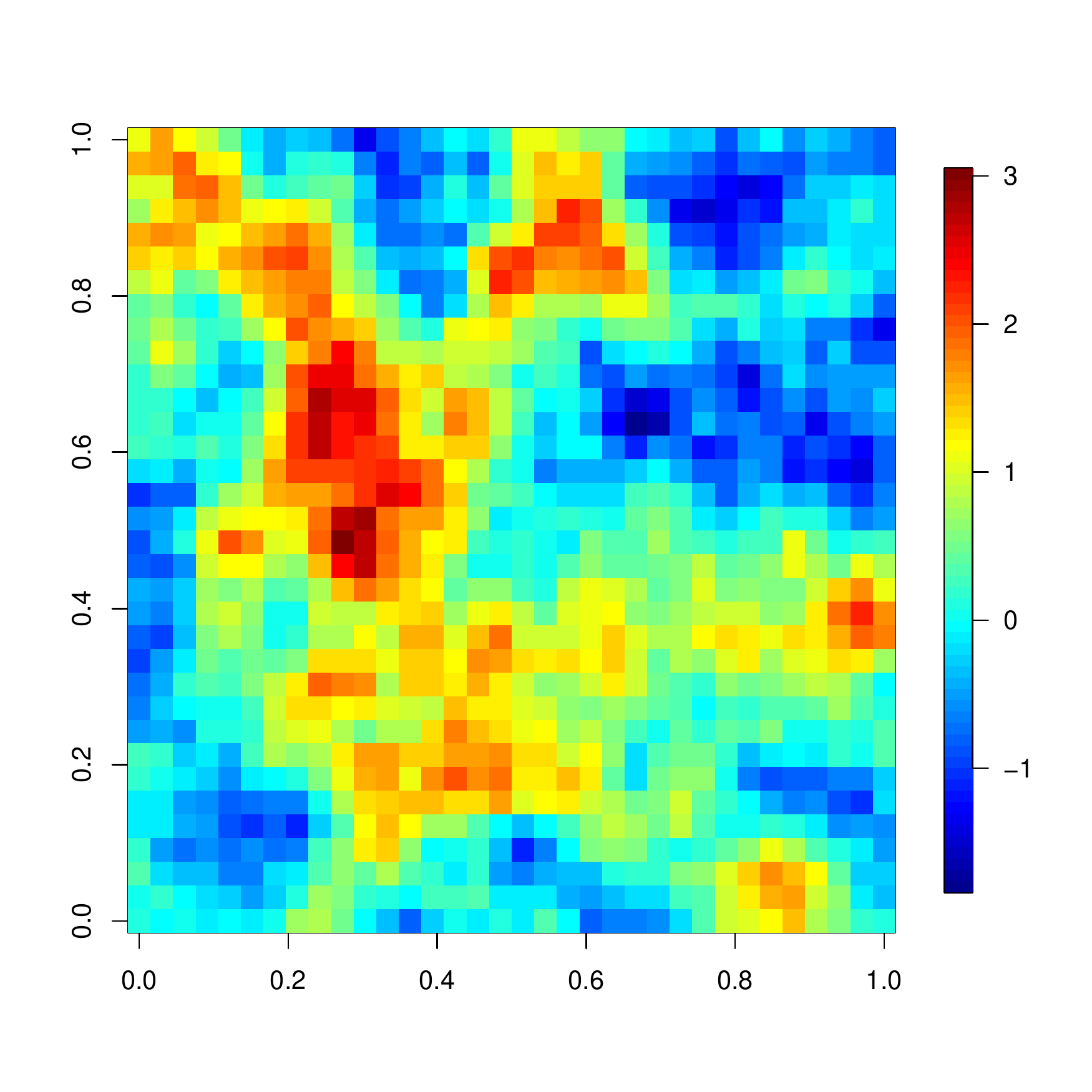}
        \caption{t=10}
    \end{subfigure}
    \begin{subfigure}[b]{0.45\textwidth}
        \includegraphics[trim={30 30 30 30},clip,width=1.0\textwidth]{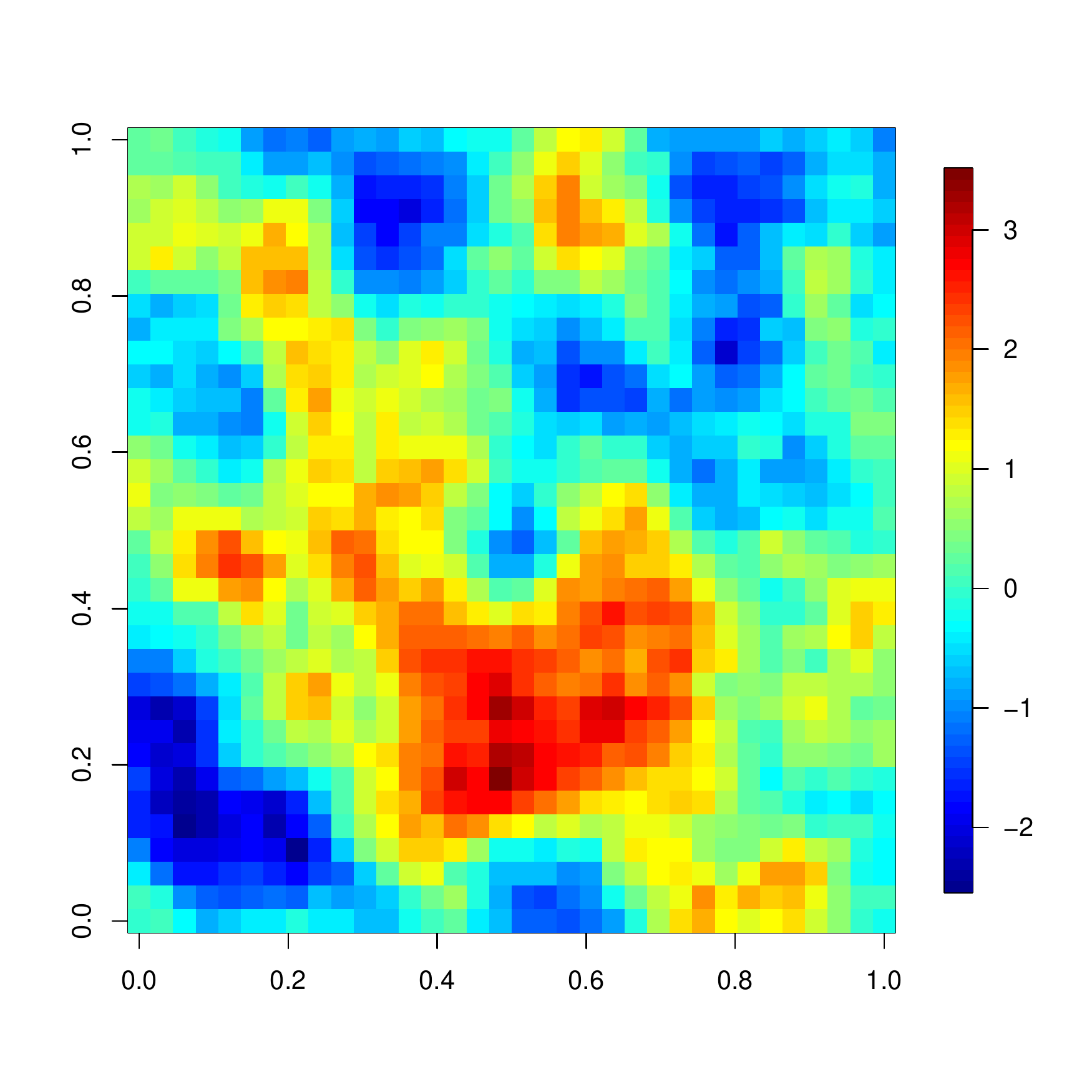}
        \caption{t=20}
    \end{subfigure}
    \caption{Sample realization of the 2D advection diffusion process described in Section \ref{sec:comparison-setup} at select time points. The first row shows two consecutive time points while the plots in the second row, corresponding to time points further apart, illustrates the long term evolution of the process.}
    \label{fig:sample}
\end{figure}

We consider an advection diffusion process $x(\bs, t)$ defined over $\mathbb{R}^2\times[0, \dots, T]$, which means that its dynamics are expressed by the following partial differential equation:
\begin{equation} \label{eq:adv-diff}
\frac{\partial x}{\partial t} = \alpha \left( \frac{ \partial^2 x }{ \partial^2 s_x } + \frac{ \partial^2 x }{ \partial^2 s_y } \right) + \beta \left( \frac{ \partial x}{\partial s_x} + \frac{\partial x}{\partial s_y} \right) + \eta,
\end{equation}
where $\eta(\bs, t)$ is a zero-mean stationary Gaussian process with an exponential covariance function with marginal variance $\sigma_{\bw}^2 = 0.1$ and range $\lambda = 0.15$. This setting of $\lambda$ allows the process to exhibit clear variation over the chosen grid (see below) but preserves substantial dependence between neighboring locations. We set $\alpha = 4\times 10^{-5}$ and $\beta = 10^{-2}$ which leads to a stable differencing scheme for our chosen grid (below) while producing visible advection and diffusion.
We also assume that $\eta(\cdot, \cdot)$ is independent across time. We then consider a regular grid of size $n_{\mathcal{G}} = 34 \times 34 = 1156$ covering the square $\mathcal{D} = [0, 1]\times [0, 1]$ and discretize $x$ over this grid using centered finite differences. 
This results in a vector $\bx_t$ with each component representing the value of $x$ at a corresponding grid point and gives a discrete version of \eqref{eq:adv-diff} which takes the form \eqref{eq:state-model}. 
We use $\bx_0 \sim \normal(\bfzero, \bfSigma_{0|0})$, where $\bfSigma_{0|0}$ corresponds to the exponential covariance function with range $\lambda$, marginal variance $\sigma^2_0 = 1$. This choice of marginal variance, 10 times greater than the marginal variance of the model error, means that most of the variation is explained by the model, but that the model error is nevertheless non-negligible. 
We further assume that at each time point $t \in [1, 2, \dots, T]$, where $T=20$, we are given a set of noisy observations $\by_t$ corresponding to some of the points from the grid. We take the measurement error to be Gaussian which means that $\by_t$ follows the data model \eqref{eq:data-model} with $\bR_t = \sigma_{\bv}^2\bI_{n_t}$
where we set $\sigma_{\bv}^2 = 0.05$ and the matrix $\bH_t$ is obtained by taking a diagonal matrix $\bI_{n_{\mathcal{G}}}$ and removing the rows which correspond to the grid points with no associated observations. This choice of $\sigma^2_{\bv}$ means that the signal to noise ratio is relatively high. A sample realization of this process at two time points is shown in Figure \ref{fig:sample}. Many other combinations of parameter values were previously considered in the case of filtering \citep{jurek2021multi}, but the relative performance of the analogues of the HV-based and low-rank filters was robust to these changes.

We then perform several numerical experiments using the following methods:
\begin{description}
\item[\textbf{scalable FFBS (Scalable)}:] Our method as described in Algorithm \ref{alg:sFFBS}.
\item[\textbf{Low-rank-based FFBS (Low-rank)}:] A sampling method based on a low-rank approximation of the latent process $x$. Within the context of our paper and for ease of comparison, we can view it as a special case of Algorithm \ref{alg:sFFBS} with the $\bS$ matrix in which only the diagonal and the first $N$ columns of $\bS$ are nonzero. This is equivalent to using the modified predictive process approach \citep[][]{Banerjee2008,Finley2009} to approximate the process $x$ and has the same computational complexity as scalable FFBS.
\item[\textbf{standard FFBS (Standard)}:] The method described in Algorithm \ref{alg:FFBS}. It can be viewed a special case of Algorithm \ref{alg:sFFBS}, in which $\bS = \mathbf{1}_{n_{\mathcal{G}}}\mathbf{1}_{n_{\mathcal{G}}}^\top$ and $N = n_{\mathcal{G}}$.
\end{description}

\subsection{Timing}

We start by showing the difference in wall-clock time required to generate a single sample using the model settings and sampling methods described in Section \ref{sec:comparison-setup}. We run our code on a high-end laptop equipped with 16 Intel i7 CPUs each with a clock speed of 2.30GHz and 16GB of memory. In order to eliminate the influence of random processes executed at the same time, we use one method at a time, measure the time elapsed from the beginning until the end of Algorithm \ref{alg:FFBS}, repeat it 10 times and report the average. The results are shown in Table \ref{tab:timing} and show that both approximate methods have a similar run time, which is much less than the run time of the standard FFBS. 
In the subsequent simulations, we show that the low-rank method, while comparable in execution time, is inferior in performance according to several criteria. 

\begin{table}[]
    \centering
    \begin{tabular}{c|c|c|c}
         method & \textbf{Standard} & \textbf{Low-rank} & \textbf{Scalable}\\
         \hline
         average time & 322.2 & 13.2 & 13.3 
    \end{tabular}
    \caption{The average time in seconds required to generate one sample from the model from Section \ref{sec:comparison-setup} using each of the sampling methods.}
    \label{tab:timing}
\end{table}

\subsection{Sampling the latent vector}\label{sec:crps}

\begin{figure}[ht]
    \centering
        \includegraphics[width=1.0\textwidth]{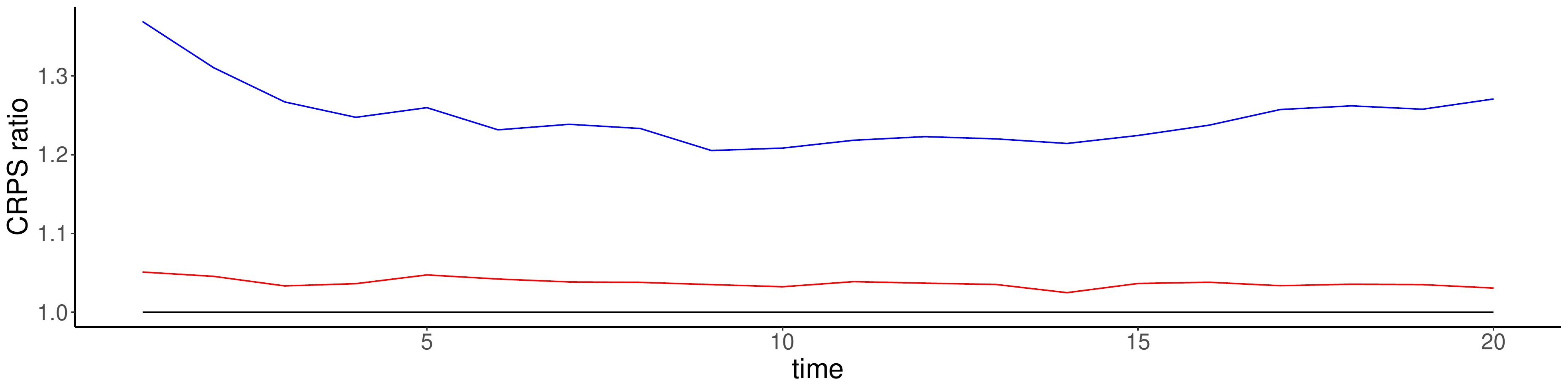}
    \caption{Average CRPS ratios for methods described in Section \ref{sec:comparison} $N_\text{samp}=50$ samples based on different versions of the HV approximation averaged over $N_\text{iter}=10$ repetitions. The red line corresponds to the scalable method, the blue line to the low-rank method and the black line to the standard sampler. All scores are calculated as relative to the score of the standard FFBS (for which its ratio with itself is therefore always equal to 1).}
    \label{fig:crps}
\end{figure}

In the second set of our simulations, we demonstrate the excellent accuracy of Algorithm \ref{alg:sFFBS} by generating a sample of size $m$ from smoothing distribution of the latent vector $\bx_t$. We then compare the results generated by other methods using continuous rank probability score (CRPS) for ensembles \citep[][Section 4.2]{Gneiting2008}. In general, if $\mathcal{Q} = \left\{q_1, \dots, q_{N_\text{samp}}\right\}$ is the ensemble of size $N_\text{samp}$ forecasting the vector $q$ we can calculate this score as
\begin{equation}
    \text{CRPS}(\mathcal{Q}, q) = \frac{1}{N_{\text{samp}}}\sum_{i=1}^{N_\text{samp}} \lVert q_i - q \rVert_2 - \frac{1}{2N_\text{samp}}\sum_{i=1}^{N_\text{samp}}\sum_{j=1}^{N_\text{samp}}\lVert q_i - q_j\rVert_2,
\end{equation}
where $\lVert\cdot\rVert_2$ denotes the second (i.e. Euclidian) norm. The lower the value of the CPRS, the more accurately the ensemble predicts the true realization $q$. Under some mild conditions, CRPS is a strictly proper scoring rule \citep{Gneiting2007,Gneiting2014}.
In order to evaluate the performance of scalable FFBS we adopt the following approach. We generate a sample of size $N_\text{samp} = 50$ using methods described in Section \ref{sec:comparison-setup} and calculate the CRPS for each of them at each time point. For each of the approximate methods we then calculate the ratio of their respective scores and the score of the standard FFBS. We repeat this procedure $N_{\text{iter}} = 10$ times and present these average score ratios in Figure \ref{fig:crps}. We conclude that scalable version of the FFBS algorithm we propose is an excellent approximation of its standard version and that it significantly outperforms the low-rank approach.




\subsection{Gibbs sampling}\label{sec:gibbs}

One of the more common applications of the standard FFBS algorithm consists in using it as one of the steps in a Gibbs sampler. In this section we demonstrate the performance of such a sampler which which relies on the methods described in Section \ref{sec:comparison-setup}

\begin{figure}[ht]
    \includegraphics[width=1.0\textwidth]{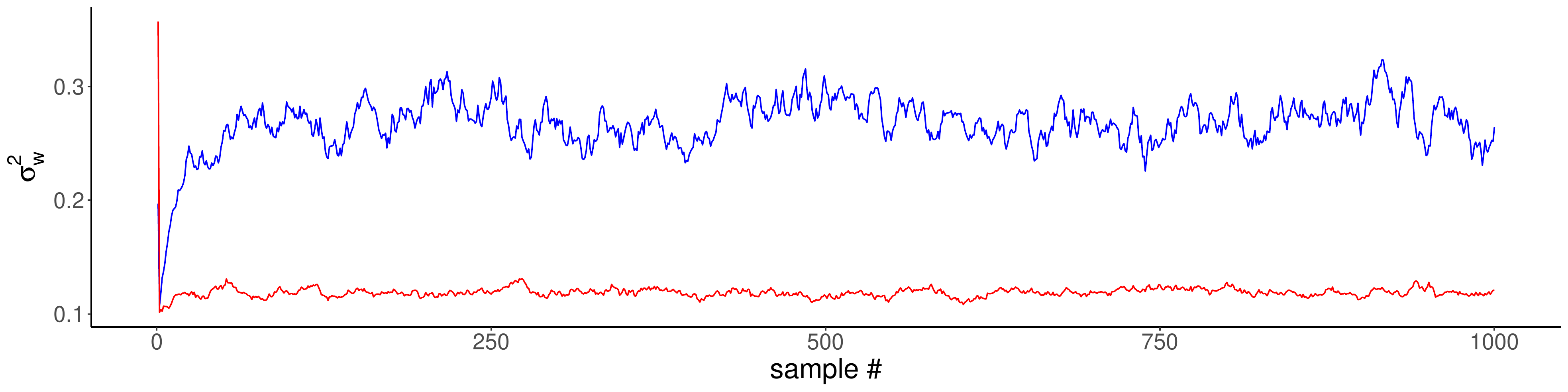}
    \caption{Samples from the conditional posterior distribution $p(\sigma_w^2|\bx_{1:T}, \by_{1:T})$. (The true value was $\sigma_{\bw}^2 = 0.1$.) The blue line denotes samples obtained using the low-rank method and the red line was generated using the scalable method. The standard method was not computationally feasible.}
    \label{fig:gibbs-sig2w}
\end{figure}

Using the model from Section \ref{sec:comparison-setup} now we assume that the $\sigma_{\bw}^2$ parameter is unknown. 
Imposing an inverse gamma prior with shape parameter $a=0.001$ and scale parameter $b=0.001$ results in the conditional posterior distribution $p(\sigma_w^2|\bx_{1:T}, \by_{1:T})$ that is inverse gamma with the shape parameter $\tilde{a} = a + \frac{n(T-1)}{2}$ and the scale parameter $\tilde{b} = b + \frac{1}{2}\sum_{t=2}^T\left( \bx_t - \bE_t\bx_{t-1}\right)^{\top} \Sigma_w^{-1} \left( \bx_t - \bE_t\bx_{t-1}\right)$. We then run the Gibbs sampler in which we sample the latent vectors $\{\bx_t\}_{t=1}^T$ using each of the methods described at the beginning of Section \ref{sec:intro} assuming that we have observations corresponding to a random selection of 30\% of grid points, i.e. $n_t = 0.3 \nG$. Figure \ref{fig:gibbs-sig2w} shows the samples from the conditional posterior distribution. In each case the sampler was initialized at a random value between 0 and 0.5.
We used only the approximate methods in the construction of the Gibbs sampler, because standard FFBS was not computationally feasible for a problem of this size. The results show that the sampler based on the scalable method generates draws of $\sigma^2_{\bw}$ that are much closer to the true value (0.1) than the draws obtained using the low-rank method.

\FloatBarrier
\section{Analysis of total precipitable water}\label{sec:tpw}

In this section we apply our proposed sampler to real observations of the amount total precipitable water (TPW) in the atmosphere, defined as the mass of the water vapor in a column of air above a given area. TPW is commonly used in numerical weather prediction, forecasting extreme weather events, or assessing fire danger in drought-stricken areas. Hence, inferring complete and noise-free spatio-temporal maps of TPW is of considerable scientific value. The collection of data we work with is comprised of the total of 47,007 measurements made over a portion of the continental United States and the Gulf of Mexico at $T=9$ points in time over a period of 40 hours in January 2011. Each data point corresponds to a cell in a $0.5^\circ \times 0.5^\circ$ latitude/longitude grid covering the area between $-125.18^\circ$W and $-107.21^\circ$W and $37.14^\circ$N and $50.07^\circ$N, resulting in 15,876 spatial grid cells. All of the observations were acquired using the Microwave Integrated Retrieval System (MIRS) satellite and are available from the authors upon request.

A superset of the data we use here has been analyzed previously using a low-rank filtering approach similar to the one described in Section \ref{sec:comparison-setup} in \citet{katzfuss2017parallel}, and a filtering approach based on the HV approximation (Algorithm \ref{alg:HVF}) in \citet{Jurek2022}. 

\begin{figure}[ht]
    \begin{subfigure}{.99\textwidth}
        \includegraphics[width=1.0\textwidth]{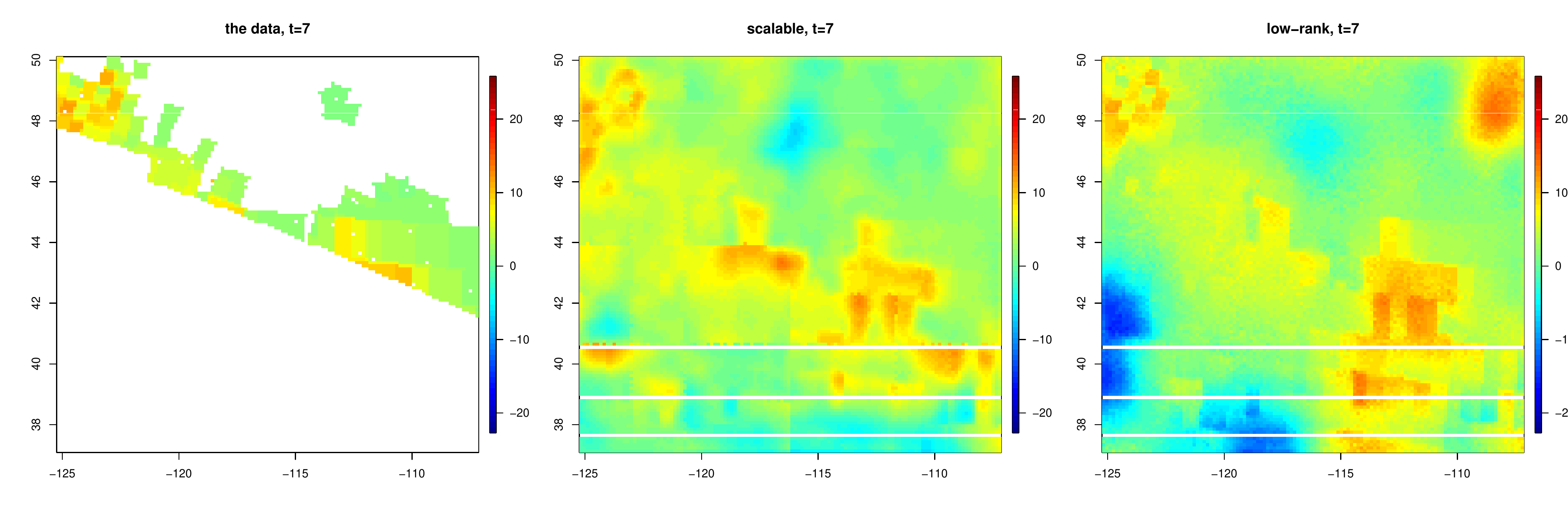}     
    \end{subfigure}
    \begin{subfigure}{.99\textwidth}
        \includegraphics[width=1.0\textwidth]{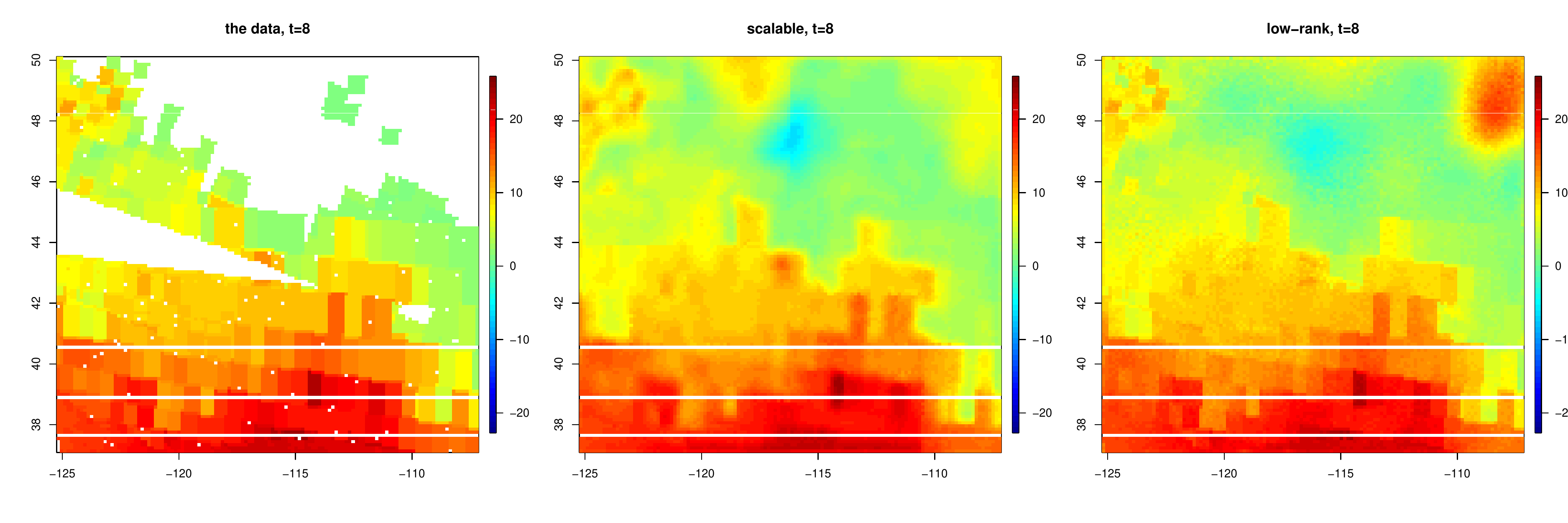}     
    \end{subfigure}
    \begin{subfigure}{.99\textwidth}
        \includegraphics[width=1.0\textwidth]{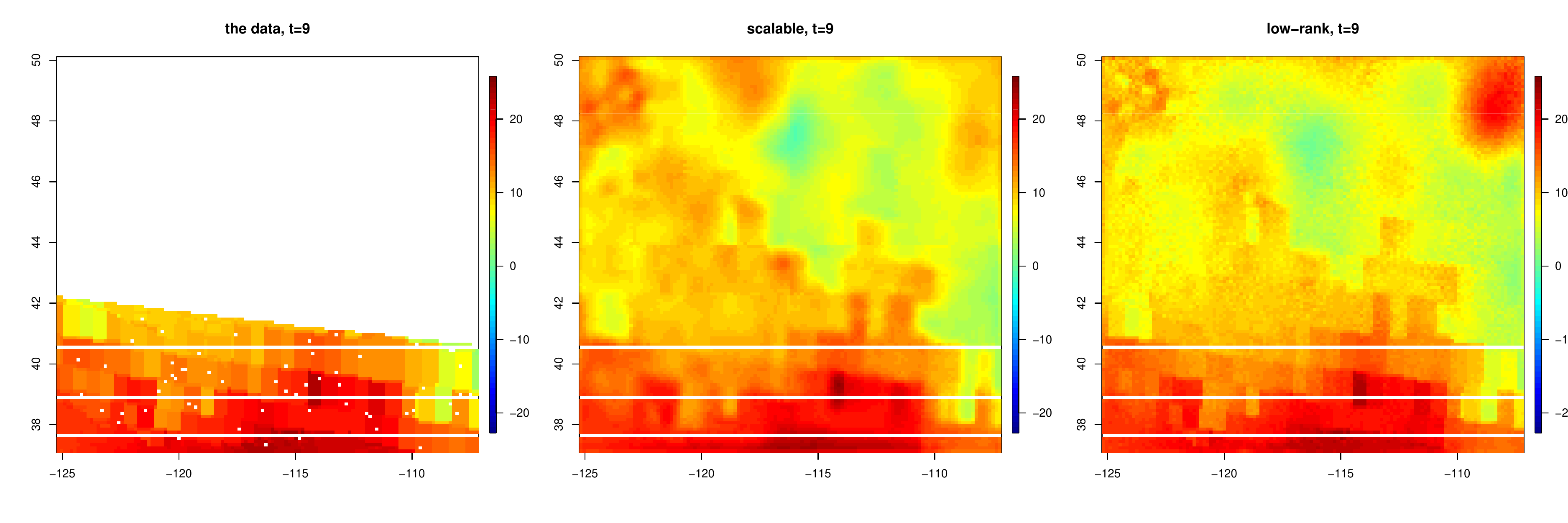}     
    \end{subfigure}
    \caption{Total precipitable water (1st column) and the pointwise mean of all the samples generated using the scalable method (2nd column) and the low-rank method (3rd column).}
    \label{fig:TPW}
\end{figure}

For a given time point $t$ we use $\by_t$ to denote the corresponding data, each of which is assumed to contain an independently and identically distributed normal measurement error with mean 0 and variance $\sigma_{\bv}^2$. At each point we calculated the mean of all measurements and subtracted it from the observations gathered at that time. The resulting value at selected time points are shown in the first column in Figure \ref{fig:TPW}. At each time point we set aside 1\% of all available measurements to be later used for result verification.

\begin{figure}[ht]
    \centering
    \includegraphics[width=1.0\textwidth]{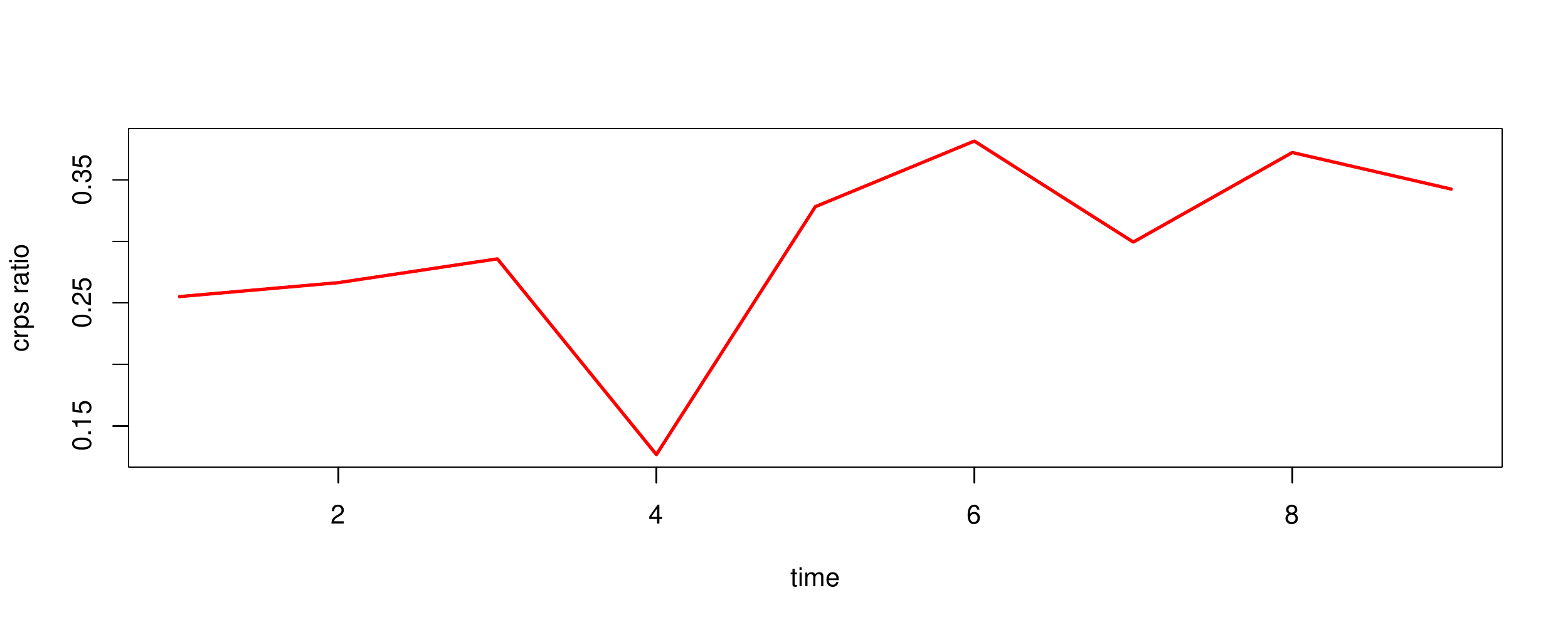}
    \caption{For the TPW data, CRPS for samples calculated using the scalable method relative to the CRPS for samples calculated using the low-rank method.}
    \label{fig:rcrps}
\end{figure}

We assume that the temporal evolution of TPW during the study period can be captured by an advection-diffusion equation, as described in Section \ref{sec:comparison-setup}. We use the diffusion coefficient to be $\alpha=0.000003$ and the advection coefficient $\beta=0$ (no advection). If $\tilde{\bE_t}$ denotes the temporal evolution operator obtained using differencing, we take $\levol_t=c\tilde{\bE_t}$ with $c=0.9$ to allow for more random variation at each time point.

We take the initial covariance function $\bfSigma_{0|0}$ to be derived from a Mat\'ern covariance function with smoothness $\nu=1.5$, range $\lambda$, marginal variance $\sigma^2_0$ and the $\bQ_t$ matrix to be derived from a Mat\'ern covariance function with the same smoothness $\nu=1.5$ and range $\lambda$ and marginal variance $\sigma^2_{\bw}$. 

In order to determine the values of parameters $\lambda, \sigma_0^2, \sigma_{\bw}^2$ and $\sigma_{\bv}^2$ we consider a purely spatial problem and assume that at each time $\bx_t$ corresponds to a discretization of a mean-zero, 2D Gaussian random field with a Mat\'ern covariance function with smoothness 1.5, range $\lambda_t$ and marginal variance $\sigma^2_t$ and that $\by_t$ are the corresponding observations with $\by_t|\bx_t = \normal(\bx_t, \tau_t^2)$. We use the Vecchia approximation with $N=80$ nonzero elements in each row of $\bS$ and use it to optimize the approximate likelihood function \citep[see][]{Zilber2021}. We take $\lambda = \frac{1}{T}\sum_t \lambda_t$, $\sigma_{\bv}^2 = \frac{1}{T}\sum_t \tau^2_t$. For the marginal variance parameters, we assume that $\sigma_0^2 + \sigma_w^2 = \frac{1}{T}\sum_t \sigma_t^2$ and then set $\sigma^2_{\bw} = (1-c) \sigma^2_0$. Table \ref{tab:TPW-params} summarizes the parameter values obtained in this way.

\begin{table}[]
    \centering
    \begin{tabular}{c|c|c|c|c|c|c}
         $\lambda$ & $\sigma_0^2$ & $\sigma_{\bw}^2$ & $\sigma_{\bv}^2$ & $c$ & $\alpha$ & $\beta$ \\
         \hline
         1.0 & 74.7 & 8.3 & 1.63 & 0.9 & 0.000003 & 0
    \end{tabular}
    \caption{Values of parameters used in the application to TPW}
    \label{tab:TPW-params}
\end{table}

Then using Algorithm \ref{alg:sFFBS} we generate $N_{\text{samp}}=20$ samples $\{\bx_{1:T}^i\}_{i=1}^{N_{\text{samp}}}$ from the smoothing distribution of the state vector $\bx_t$ using the scalable method and the low rank method with the conditioning set of size $N=52$. 
In Figure \ref{fig:TPW} we present the mean field $\bar{\bx} = \frac{1}{N_{\text{samp}}}\sum_{i=1}^{N_{\text{samp}}}\bx_t^i$ for select values of $t$.

In order to evaluate our method we use CRPS as described in Section \ref{sec:crps} using the observations which we set aside at the beginning. Because of the scale of the problem, the standard method was not feasible. Instead we report the ratio $\text{rCRPS} = 100\%  \frac{\text{CRPS}_{\text{HV}}}{\text{CRPS}_{\text{LR}}}$, which tells us by what percentage the score is reduced, if we use the scalable method as opposed to the low-rank method. We report the rCRPS for each time point in Figure \ref{fig:rcrps}. The results show, that using the scalable method instead of the low-rank method leads to about 20\% lower CRPS at a typical point in time.

\section{Conclusions\label{sec:conclusion}}

Our paper proposes an approximate method of sampling the latent state in the context of linear Gaussian state space models. Our approach, called scalable FFBS, can be applied even to fields with tens of thousands of random variables. It also outperforms samplers based on a popular low-rank approximation according to several important metrics. 
The proposed algorithm can be extended in several directions. First, combining it with the Laplace approximation, similar to \citep{Jurek2022}, it can be applied to a large class of non-Gaussian distributions. Using the correlation distance \citep{kang2021correlation}, might allow to accommodate data without a clear spatial structure. We also envision extending our framework to incorporate several random fields and non-linear temporal evolution.

\FloatBarrier
\footnotesize
\appendix
\section*{Acknowledgments}

MK's research was partially supported by NSF Grants DMS--1654083, DMS--1953005, and CCF--1934904, and by the National Aeronautics and Space Administration (80NM0018F0527). We would like to thank Kate Calder, Mevin Hooten, and Cory Zigler for helpful comments and discussions. Special thanks to Pulong Ma, who first pointed out the possibility of using the Vecchia approximation in the context of an FFBS algorithm, and to Dorit Hammerling for helping us access the TPW data.


\bibliographystyle{apalike}
\bibliography{references}

\end{document}